\documentclass[prl,aps,twocolumn,showpacs]{revtex4}
\usepackage{graphicx}
\usepackage{amsmath}
\usepackage{bm}
\usepackage{amstext}
\usepackage{amsxtra}

\usepackage{times}

\begin{document}

\newcommand{\To}{T_c^0}
\newcommand{\kB}{k_{\rm B}}
\newcommand{\dT}{\Delta T_c}
\newcommand{\lo}{\lambda_0}
\newcommand{\cs}{$\clubsuit$}
\newcommand{\thold}{t_{\rm hold}}

\title{Effects of Interactions on the Critical Temperature of a Trapped Bose Gas}

\author{Robert P. Smith, Robert L. D. Campbell, Naaman Tammuz, and Zoran Hadzibabic}
\affiliation{Cavendish Laboratory, University of Cambridge, J.~J.~Thomson Ave., Cambridge CB3~0HE, U.K.}

\begin{abstract}
We perform high-precision measurements of the condensation temperature of a harmonically-trapped atomic Bose gas with widely-tuneable interactions.
For weak interactions we observe a negative shift of the critical temperature in excellent agreement with mean-field theory. However for sufficiently strong interactions we clearly observe an additional positive shift, characteristic of beyond-mean-field critical correlations. We also discuss non-equilibrium effects on the apparent critical temperature for both very weak and very strong interactions.

\end{abstract}

\date{\today}



\maketitle

The effect of inter-particle interactions on the Bose-Einstein condensation temperature of a dilute gas has been a topic of theoretical debate for more than fifty years, since the pioneering work of Lee and Yang \cite{Lee:1957TcShift, Lee:1958TcShift2}. 
In a uniform system there is no interaction shift of the critical temperature $T_c$ at the level of mean-field theory. However, consideration of the correlations between particles which develop near the critical point leads to the conclusion that repulsive interactions \textit{enhance} condensation, i.e. shift the condensation temperature above the ideal gas value $\To$
\cite{Bijlsma:1996, Baym:1999, Holzmann:1999, Reppy:2000, Arnold:2001, Kashurnikov:2001, Holzmann:2001, Baym:2001, Andersen:2004RMP, Holzmann:2004}.

Ultracold atomic gases offer an excellent testbed for fundamental theories of Bose-Einstein condensation and related many-body phenomena \cite{Dalfovo:1999, Bloch:2008}. However, in these systems the problem of the interaction shift of $T_c$ is even more complex because they are produced in harmonic traps. In this case, at least for weak interactions, the $T_c$ shift is dominated by an opposing mean-field effect, which reduces the critical temperature \cite{Giorgini:1996}. 
Within experimental precision, previous measurements 
\cite{Ensher:1996, Gerbier:2004,Meppelink:2010} were consistent with the mean-field theory and could not discern the effects of critical correlations.

In this Letter, we report on high-precision measurements of the $T_c$ shift in a 
potassium ($^{39}$K) gas with tuneable interactions \cite{Roati:2007,Campbell:2010}. We employ a Feshbach resonance \cite{Chin:2010} to extend the previously explored range of interaction strengths and eliminate several key sources of statistical and systematic errors. This allows us to clearly reveal the long-sought beyond-mean-field effects on the critical temperature. We also examine the stringent requirements for equilibrium $T_c$ measurements,
which are violated 
in the regimes of 
either very weak or very strong interactions. 
In non-equilibrium gases we observe evidence for `super-heated' condensates which survive at an apparent temperature above the equilibrium $T_c$, suggesting that strong dissipation can stabilise the coherent condensed state.

Historically, most theoretical work focused on a spatially uniform gas, 
and for several decades there was no consensus on the functional form, or even on the sign of the $T_c$ shift (for an overview see {\it e.g.} \cite{Arnold:2001, Baym:2001, Andersen:2004RMP, Holzmann:2004}). It is now generally believed that the shift is positive and to leading order 
given by \cite{Arnold:2001, Kashurnikov:2001}:
\begin{equation}
\frac{\dT}{\To}  \approx 1.3 \, a n^{1/3} \approx 1.8  \, \frac{a}{\lo} \, ,
\label{eq:uniformTcShift}
\end{equation}
where $\dT = T_c - \To$, $a >0$ is the s-wave scattering length,
$n$ the particle density, and $\lo$ the thermal wavelength at temperature $T_c^0$. The positive $\dT$  implies that condensation occurs at a phase space density below the ideal gas critical value of $n \lambda^3 = \zeta(3/2) \approx 2.612$ (where $\zeta$ is the Riemann function).

\begin{figure}[h]
\centering
\includegraphics[width=0.95\columnwidth]{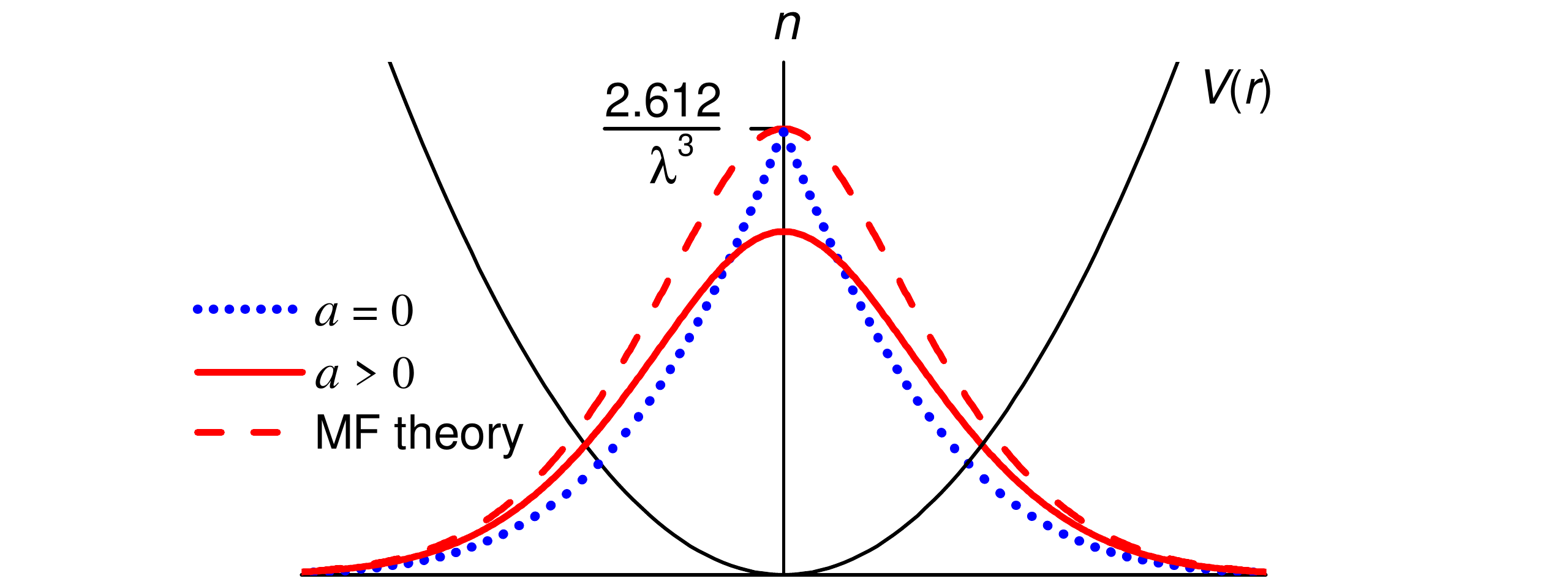}
\caption{(Color online) Opposing effects of interactions on the critical point of a Bose gas in a harmonic potential $V(r)$. Compared to an ideal gas (dotted blue line) with the same $T_c$, repulsive interactions reduce the critical density, but also broaden the density distribution (solid red line). Mean-field theory (dashed line) captures only the latter effect, and predicts an increase of the critical atom number $N_c$ at fixed temperature $T$, equivalent to a decrease of $T_c$ at fixed $N$.}
\label{fig:TcCartoon}
\end{figure}

\begin{figure} [t]
\centering
\includegraphics[width=\columnwidth]{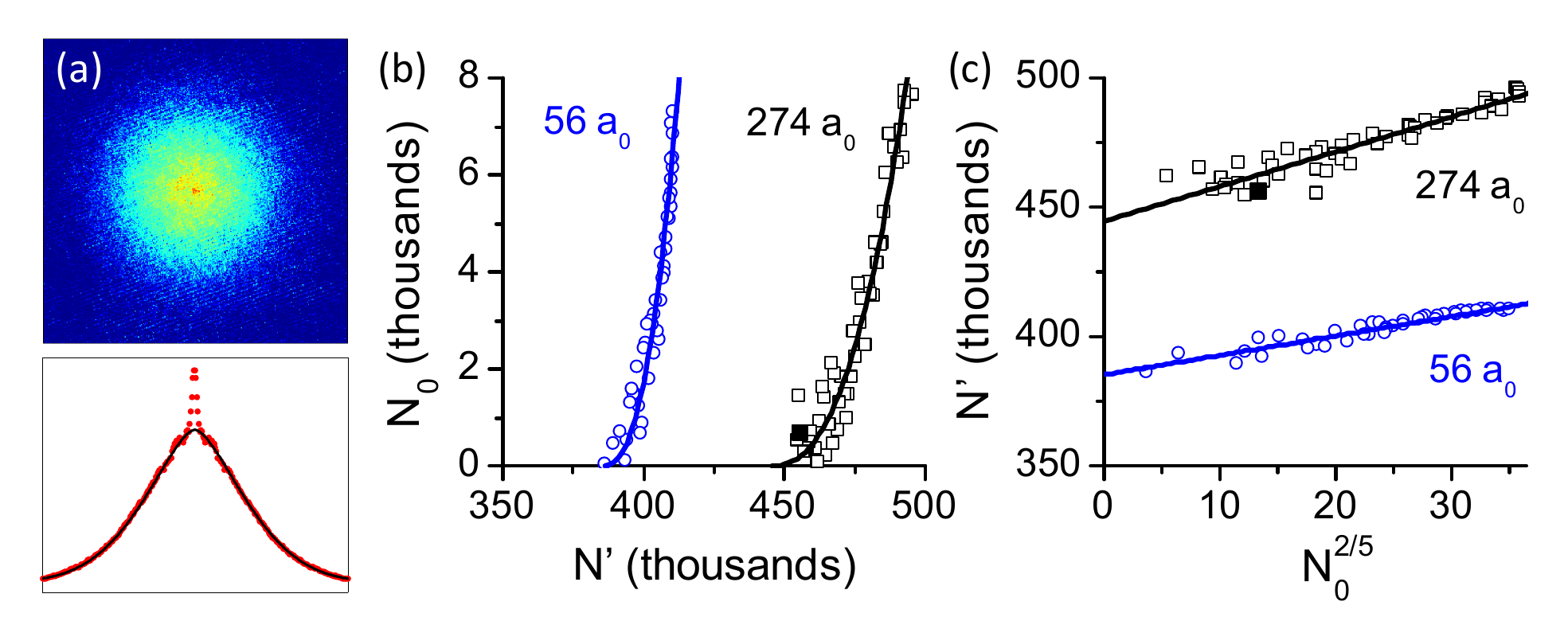}
\caption{(Color online) Determination of the critical point and differential interaction shift. (a) An absorption image of a cloud with 0.14\,\% condensed fraction, and the corresponding (azimuthally averaged) cut through the column density. The gas was prepared at a large scattering length, $a= 274\, a_0$, but the interactions were turned off in TOF. (b) Condensed ($N_0$) versus thermal ($N'$) atom number for two concurrently taken series with $a = 56\,a_0$ (blue circles) and $a=274\,a_0$ (black squares). Note that all points correspond to  condensed fractions below $2\,\%$. The data is scaled to the same temperature ($T=240\,$nK) and shows the shift of the critical point in the form $\Delta N_c(T)$. The solid point corresponds to the image shown in (a). Solid lines show the extrapolation to $N_0=0$, necessary  to accurately determine $N_c$. (c) $N'$ is plotted versus $N_0^{2/5}$ for the same data as in (b), showing more clearly the extrapolation procedure.}
\label{fig:Nc}
\end{figure}

For a harmonically trapped gas, the condensation temperature is 
defined for  a given atom number $N$, rather than for a given density $n$. For an ideal gas, $\kB \To = \hbar \bar{\omega} \, \left[ N/\zeta(3) \right]^{1/3}$,
where $\bar{\omega}$ is the geometric mean of the trapping frequencies along the three spatial dimensions, and $\zeta(3) \approx 1.202$. 
This corresponds to a phase space density in the centre of the trap equal to the uniform system critical value, $n(0)\lambda^3 = \zeta(3/2)$. The interaction shift of the critical point can be expressed either as $\Delta T_c(N)$ (for comparison with theoretical literature) or as $\Delta N_c(T)$ (for easier visualisation, as in Figs.~\ref{fig:TcCartoon} and \ref{fig:Nc}).

The two opposing effects of repulsive interactions on the critical point of a trapped gas are illustrated in Fig.~\ref{fig:TcCartoon}, where we sketch the density distribution at the condensation point for an ideal (dotted blue line) and an interacting (solid red line) gas at the same temperature. In the spirit of the local density approximation, the critical density should be reduced by repulsive interactions. However, interactions also broaden the density distribution.   For weak interactions the latter effect is dominant, making the overall interaction shift $\Delta N_c(T)$ positive, or equivalently $\Delta T_c(N)$ negative.

The negative $T_c$ shift due to the broadening of the density distribution in a harmonically trapped gas can be calculated using mean-field (MF) theory, which neglects the reduction of the critical phase space density implied by Eq.~(\ref{eq:uniformTcShift}) (see dashed line in Fig.~\ref{fig:TcCartoon}).
This approach gives \cite{Giorgini:1996}:
\begin{equation}
\frac{\dT}{\To} \approx - 3.426 \, \frac{a}{\lo} \, .
\label{eq:StringariMF}
\end{equation}
The dominance of the negative MF shift of $T_c$ over the positive beyond-MF one goes beyond the difference in numerical pre-factors in Eqs.~(\ref{eq:StringariMF}) and (\ref{eq:uniformTcShift}). At the condensation point, in a non-uniform system only the central region of the cloud is close to criticality, which reduces the net effect of critical correlations so that they are expected to affect $T_c$ only at a higher order in $a/\lo$. The MF result of Eq.~(\ref{eq:StringariMF}) should therefore be exact at first order in $a/\lo$.
There have been several attempts to theoretically combine the effects of MF repulsion and beyond-MF correlations on  $T_c$ for a harmonically trapped gas \cite{Houbiers:1997, Holzmann:1999b, Arnold:2001b, Davis:2006, Zobay:2009}, but no consensus has been reached 
beyond the expectation that the additional beyond-MF shift should be positive.

Previous measurements \cite{Ensher:1996, Gerbier:2004,Meppelink:2010}, consistent with Eq.~(\ref{eq:StringariMF}) within experimental errors, were performed for $a/\lo $ ranging from $0.007$ \cite{Meppelink:2010} to  0.024 \cite{Gerbier:2004}. We explore the range $0.001 < a/\lo < 0.06$, using the 402.5 G Feshbach resonance in the $|F,m_F\rangle = |1,1\rangle$ lower hyperfine state of $^{39}$K \cite{Zaccanti:2009}. As in \cite{Campbell:2010, Tammuz:2011}, we produce  $^{39}$K condensates in a crossed optical dipole trap which provides a close to isotropic trapping potential. Near the bottom of the trap, $\bar{\omega}/2\pi =75- 85\,$Hz for our measurements with atom numbers $N \approx (2-8) \times 10^5$.

To measure the critical point we prepare a partially condensed cloud, fix the optical trap depth, and let the atom number decay towards $N_c$ 
through
inelastic processes; 
meanwhile, elastic collisions redistribute particles between condensed and thermal components, and the temperature remains essentially constant \cite{Tammuz:2011}. Initial preparation of clouds with various condensed fractions is done at $a = 135 \, a_0$, where $a_0$ is the Bohr radius. We then adjust $a$ to the desired value by ramping the Feshbach field, and wait for an $a$-dependent hold time $t_{\rm hold}$ before releasing the gas from the trap and measuring its momentum distribution through absorption imaging after 19 ms of time-of-flight (TOF). 
In the last part of the paper we discuss the strict requirements on the relationship between $\thold$, the elastic scattering rate $\gamma_{\rm el}$, and the relevant  atom-number decay time $\tau$ for the measurements of $T_c$ to faithfully reflect equilibrium properties of the gas. For now we focus on presenting the measurements which we trust to be in equilibrium.

In addition to extending the $a/\lo$ range,  the Feshbach resonance provides us with two experimental advantages essential for both the precision and accuracy of our $\Delta T_c$ measurements:

(1) 
For each measurement series at a given $a$ and $\lo$, we concurrently take a reference measurement with a different $a$, same $\bar{\omega}$ and very similar $N$, hence very similar $\lo$. Specifiically, for the reference point we choose a small $a$ such that $a/\lo \approx 0.005$.
We thus directly access the small $T_c$ shift due to the difference in $a/\lo$, and essentially eliminate all $a$-independent systematic errors that usually affect absolute measurements of $T_c(N, \bar{\omega}, a)$. These include uncertainties in the absolute calibration of $N$ and $\bar{\omega}$, as well as the additional $T_c$ shifts due to finite-size effects \cite{Dalfovo:1999} and the small anharmonicity of the trapping potential \cite{Campbell:2010}. 

(2) We home in on the critical point by turning off the interactions during TOF. To do this we quickly (in $\lesssim 2\,$ms) ramp the Feshbach field to $350\,$G immediately after the release of the gas from the trap. This minimises the expansion of small condensates and allows us to reliably detect condensed fractions as small as  $\sim 10^{-3}$ (see Fig.~\ref{fig:Nc}).

Fig.~\ref{fig:Nc} illustrates our differential measurement. Here $a=274\,a_0$, $\lambda_0 \approx 10^4\,a_0$, and $a = 56\,a_0$ for the reference series.
If the two series had identical $N_c$ values, we could directly read off the differential $\Delta T_c(N)$. To correct for the small (few \%) difference in $N_c$ we apply the ideal gas scaling, $T_c \propto N^{1/3}$, to the reference series. The second-order error in $\Delta T_c$ due to the small ($ < 2 \,\%$) $T_c$ shift at $a/\lo \approx 0.005$ is much smaller than our statistical error bars. For visual clarity, in Fig.~\ref{fig:Nc} we instead scale to the same temperature and display $\Delta N_c(T)$. 

In Fig.~\ref{fig:Nc}(b-c) we show the relationship between the condensed ($N_0$) and thermal ($N'$) atom number near the critical point~\cite{smallNo}. 
The rise of $N_0$ in Fig.~\ref{fig:Nc}(b) is not simply vertical because the thermal component in a partially condensed gas is not saturated at $N_c$ \cite{Tammuz:2011}; one can also see that this effect is more pronounced at higher $a$. 
It is therefore essential to carefully extrapolate $N'$ to the $N_0=0$ limit in order to accurately determine $N_c$. 
We extrapolate using 
$N'=N_c+S_0 N_0^{2/5}$, with the non-saturation slope $S_0(T,\bar{\omega},a)$ calculated with no free parameters following \cite{Tammuz:2011}.

\begin{figure} [h]
\includegraphics[width=0.98\columnwidth]{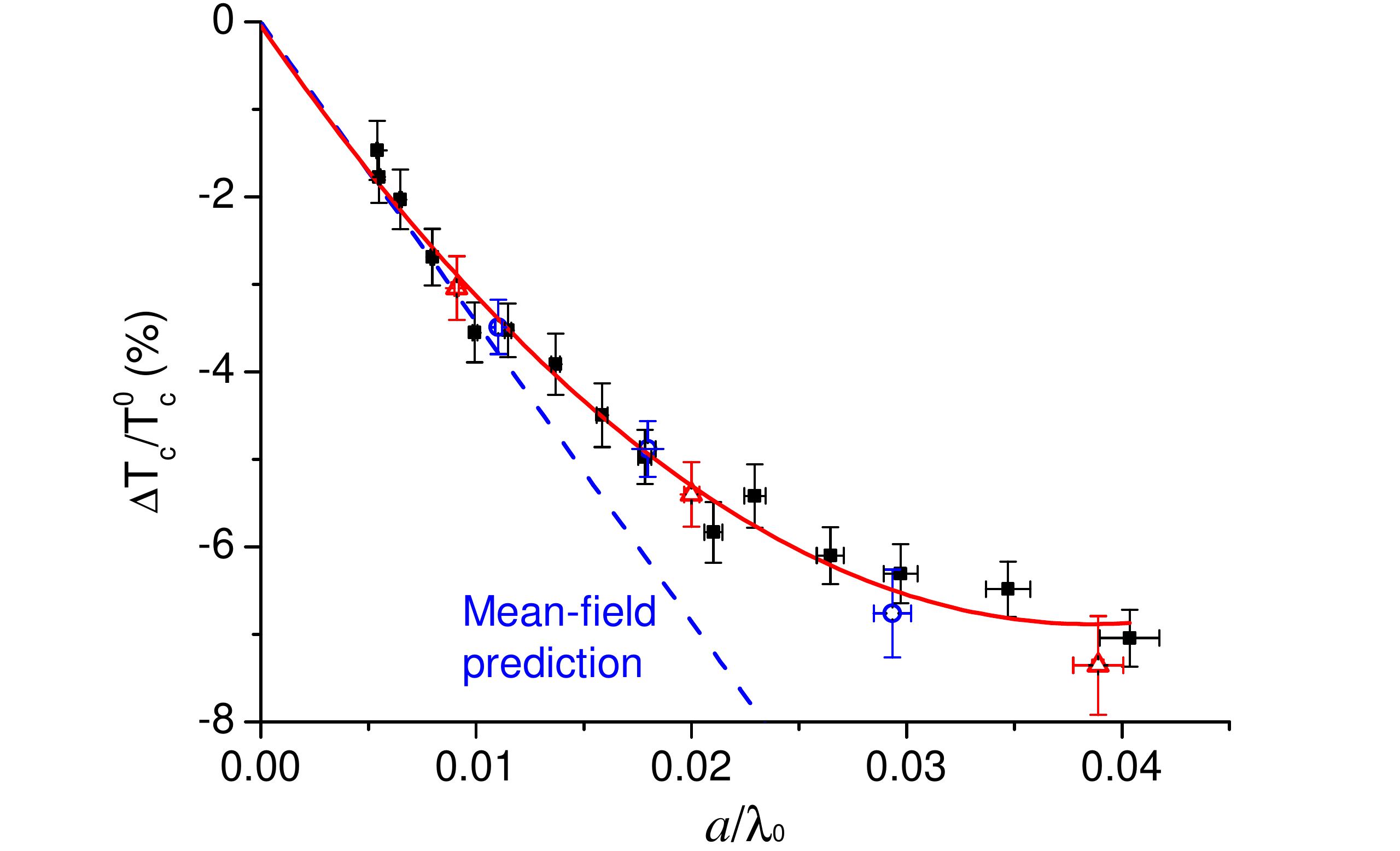}
\caption{(Color online) Interaction shift of the critical temperature. Data points were taken with $N \approx 2 \times 10^5$ (blue circles), $4 \times 10^5$ (black squares), and $8 \times 10^5$ (red triangles) atoms. The dashed line is the  
mean-field result $\Delta T_c / \To = -3.426 \, a/\lo$. The solid line shows a second-order polynomial fit to the data (see text). Vertical error bars show standard statistical errors. Horizontal error bars reflect the $0.1\,$G uncertainty in the position of the Feshbach resonance.}
\label{fig:TcShift}
\end{figure}

In Fig.~\ref{fig:TcShift} we summarize our equilibrium measurements of the interaction shift $\Delta T_c / \To$. 
We took data with a range of atom numbers, $N \approx (2-8) \times 10^5$ (corresponding to $\To \approx 180-330\,$nK), in order to verify that our results depend only on the interaction parameter $a/\lo$. 
The dashed blue line shows the MF result of Eq.~(\ref{eq:StringariMF}), which agrees very well with the data for $a/\lo \lesssim 0.01$. For larger $a/\lo$ we observe a clear deviation from this prediction. All data points are fitted well by a second-order polynomial (solid red line), $\Delta T_c/\To = b_1(a/\lo)+b_2(a/\lo)^2$, with $b_1= - 3.5 \pm 0.3$ and $b_2 = 46 \pm 5$ \cite{absolute}. From a theoretical point of view this functional form might not be exact, for example one might also expect small logarithmic corrections (see e.g. \cite{Arnold:2001b}). However within our error bars such corrections are not discernible. 

The value of $b_1$ is in excellent agreement with the MF prediction of $-3.426$ \cite{Giorgini:1996}.  The value of $b_2$ strongly excludes zero, and its sign is consistent with the expected effect of beyond-MF critical correlations. These measurements provide the first clear observation of beyond-MF effects on the transition temperature of a harmonically trapped gas.

To conclude this part of the paper, we assess the systematic errors in our measurements. In general interactions increase the kinetic energy of thermal atoms during TOF; this results in an $a$-dependent error in $T$ which does not cancel out in our differential measurements.
This error is minimized by fitting the high-energy wings of the thermal distribution (excluding the central thermal radius from the fit) \cite{Gerbier:2004c}. We also turn the interactions off at the beginning of TOF, but the reduction of $a$ is gradual over $\approx 2\,$ms. We measured the difference between (apparent) $T$ with interactions `on' and `off' during TOF to be approximately linear in $a/\lo$, and about $4\,\%$ for $a=400\,a_0$ and $\lo \approx 10^4\,a_0$. 
By varying the time at which we turn off $a$, we estimate our residual error to be $1 - 2\,\%$ at $a/\lo =0.04$. This estimate is supported by numerical simulations which reproduce the experiments quite well.
Additionally, interactions modify the initial in-trap momentum distribution. This {\em reduces} the apparent $T$ because the positive chemical potential preferentially enhances population of low-energy states. 
We numerically estimate this effect to be also approximately linear in $a/\lo$, and about $ - 2\,\%$ at $a/\lo =0.04$. 
Fortuitously, the in-trap and in-TOF effects partially cancel, resulting in a net error in $\Delta T_c/\To$ of at most $\pm 1\,\%$ at $a/\lo =0.04$.


In the rest of the paper we discuss the equilibrium conditions required for our measurements, and the non-equilibrium effects revealed when they are violated.

In general, a system with continuous dissipation can only be `close to' thermodynamic equilibrium. For an atomic gas, the proximity to equilibrium depends on the dimensionless parameter $\gamma_{\rm el} \tau$, which measures the relative rates of elastic and inelastic processes.
In practice the  $\gamma_{\rm el}$ required for equilibrium measurements also depends on the measurement precision.
We measure $N_c$ to about $1\,\%$, so we require that the gas continuously (re-)equilibrates on a timescale $\tau$ corresponding to only $1\,\%$ atom-loss. We thus require about 100 times higher $\gamma_{\rm el}$ than one would naively conclude by taking the $1/e$ lifetime of the cloud as the relevant timescale.
Equilibration is usually considered to take about $3$ collisions per particle \cite{Monroe:1993};
for all measurements shown in Fig.~\ref{fig:TcShift} we made sure that $\gamma_{\rm el} \tau > 5$. All our data also satisfy the condition $\thold > \tau > 1/\bar{\omega}$, necessary for global equilibrium to be established.

\begin{figure}
\vspace{2mm}
\includegraphics[width=0.98\columnwidth]{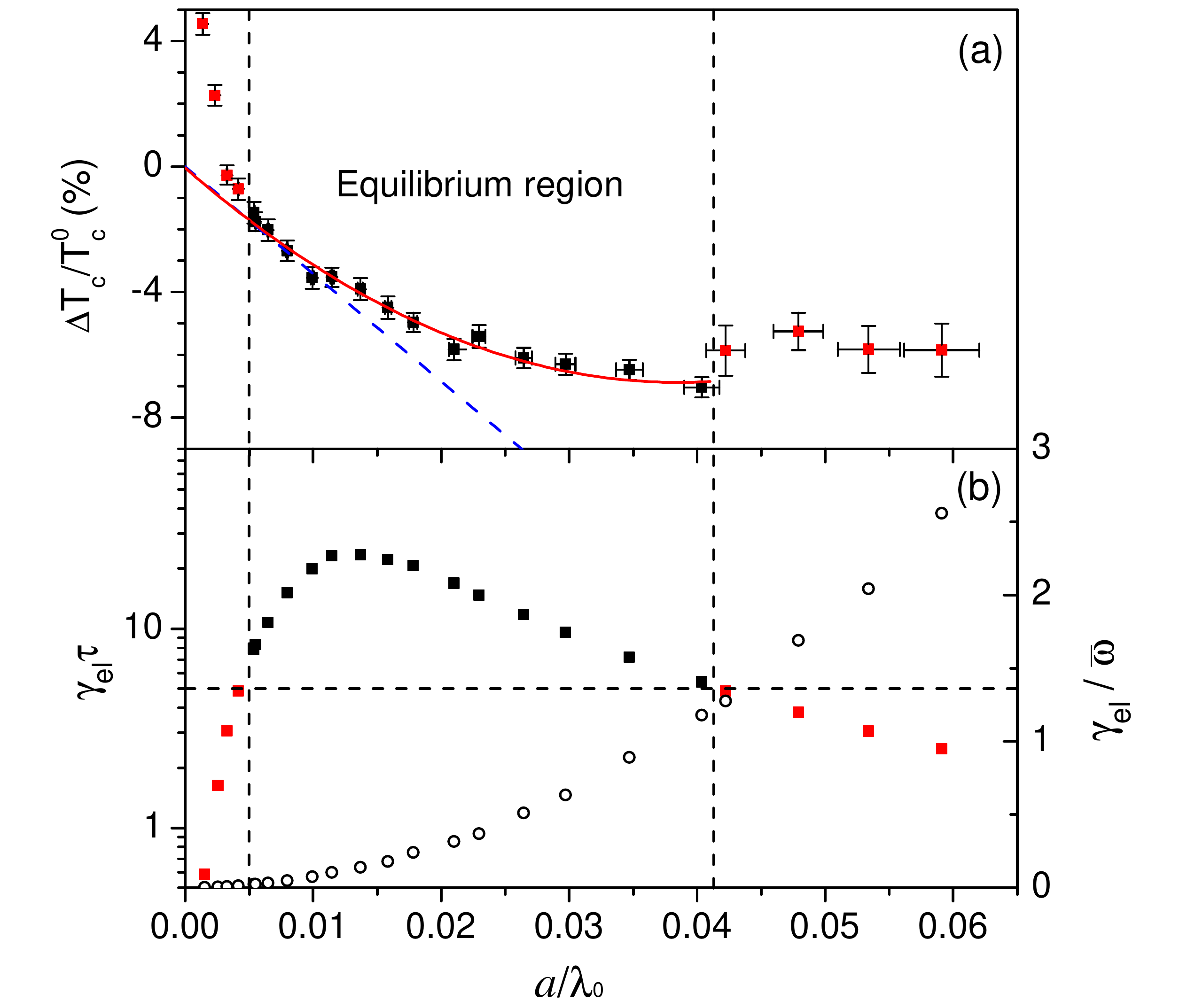}
\caption{(Color online) Non-equilibrium effects. 
(a) $\Delta T_c / \To$ for $N \approx 4 \times 10^5$ atoms is determined following the procedure which assumes equilibrium (as in Fig.~\ref{fig:Nc}). At both very low and very high $a$ the apparent $T_c$ deviates from the equilibrium curve. (b)  
Equilibrium criteria (see text):
$\gamma_{\rm el} \tau$ (solid squares) is the number of elastic collisions per particle during $1\,\%$ atom-loss; $\gamma_{\rm el}/\bar{\omega} = 1$ (open circles) marks the onset of the hydrodynamic regime.} 
\label{fig:NonEq}
\end{figure}

An interesting question in its own right is what happens if we violate these stringent equilibrium criteria. 
In Fig.~\ref{fig:NonEq}(a) we show measurements with $N \approx 4 \times 10^5$ atoms ($\lo \approx 10^4\,a_0$), extending beyond the equilibrium region shown in Fig.~\ref{fig:TcShift}. We still show only measurements that satisfy $\thold > \tau > 1/\bar{\omega}$ and $\gamma_{\rm el} \thold > 5$, so that there is nominally enough time for global equilibrium to be established. However if $\gamma_{\rm el} \tau$ is not large enough, the elastic collisions cannot `keep up' with the continuously present dissipation.
The resulting non-equilibrium effects can thus not be eliminated by simply extending $\thold$, but are an intrinsic property of the system. 
In Fig.~\ref{fig:NonEq}(b) we plot $\gamma_{\rm el} \tau$, based on calculated $\gamma_{\rm el}$ \cite{gamma} and $\tau$ measured near the critical point. Individually, $\gamma_{\rm el} \approx 0.7 - 1000\,{\rm s}^{-1}$ and $\tau \approx 2\,{\rm ms} - 1\,{\rm s}$ vary vastly as a function of $a$ ($\gamma_{\rm el}$ increasing and $\tau$ decreasing), but the breakdown of equilibrium appears to occur at very similar values of $\gamma_{\rm el} \tau$ in the low- and high-$a$ limit. 

Non-equilibrium phenomena necessarily depend on additional factors such as the initial conditions, and we therefore do not expect our quantitative results to be universal. The details of non-equilibrium dynamics will be a subject of future experiments, and here we discuss only qualitative trends. 

In the small-$a$ limit we observe a smooth rapid rise of the apparent $T_c$ above the equilibrium curve (and hence above $\To$ for $a \rightarrow 0$).
We can qualitatively understand this effect within a simple picture. In this regime, losses are most likely dominated by one-body processes which equally affect $N_0$ and $N'$. The net effect of equilibrating elastic collisions would therefore be to transfer atoms from the condensate to the thermal cloud. 
However the dissipation rate is too high compared to $\gamma_{\rm el}$, and so 
$N_0$ remains non-zero even after the total atom number drops below the equilibrium critical value $N_c$ (i.e. the measured $T_c$ is above the equilibrium value).
Note that strictly speaking $T$ is not defined out of equilibrium, but the absolute value of the observed effect is sufficiently small that an equilibrium distribution function fits the data very well and provides a good measure of the energy content of the cloud. 

Our measurements in the large-$a$ limit suggest that the initial breakdown of equilibrium again results in condensates surviving above the equilibrium $T_c$. 
However the physics in this regime is much richer, with several potentially competing effects requiring further investigation. For example, three-body decay affects $N_0$ and $N'$ differently, the thermal component is far from saturation \cite{Tammuz:2011}, and the gas also enters the hydrodynamic regime, $\gamma_{\rm el}/\bar{\omega} > 1$ [see Fig.~\ref{fig:NonEq}(b)]. 

In conclusion, we have performed high-precision studies of the effects of interactions on Bose-Einstein condensation of a trapped atomic gas. In the regime where equilibrium measurements are possible, our most important observation is the clear deviation from mean-field behaviour for sufficiently strong interactions. The additional positive shift of the critical temperature is a clear signature of the condensation-enhancing effect of critical fluctuations. These measurements should provide motivation and guidance for further theoretical studies of this difficult problem. 
We also studied non-equilibrium condensation phenomena, for both very weak and very strong interactions. Further study of these effects should prove useful for understanding condensation in intrinsically out-of-equilibrium systems, such as polariton gases.

We thank M. Holzmann and J. Dalibard for useful discussions, and M. K\"ohl for comments on the manuscript. This work was supported by EPSRC (Grant No. EP/I010580/1). 
R.P.S. acknowledges support from the Newton Trust.


\begin{thebibliography}{99}
\bibitem{absolute}
The absolute calibration of the $y$-axis in Fig.~\ref{fig:TcShift} is done by ensuring that this fit extrapolates to $\Delta T_c / \To =0$ for $a/\lo \rightarrow 0$.
If we instead simply assume that Eq.~(\ref{eq:StringariMF}) is exact for $a/\lo = 0.005$, all our results remain the same within error bars.
\end{thebibliography}

\end{document}